\documentclass[pra,showpacs,twocolumn,longbibliography,secnumarabic]{revtex4-2}
\usepackage{graphicx}
\usepackage{amssymb}
\usepackage{float}
\usepackage[T1]{fontenc}
\usepackage{amsmath}
\usepackage{mathtools}

\begin{document}
\title
{Energy of the symmetrization entanglement}
\author{Mehmet Emre Tasgin$^{*}$}
%\affiliation{$^\dagger$ Contributed equally}
\affiliation{$^*$ correspondence: metasgin@hacettepe.edu.tr and metasgin@gmail.com}
\affiliation{Institute  of  Nuclear  Sciences, Hacettepe University, 06800 Ankara, Turkey}
 \date{\today}
%\onecolumngrid
\begin{abstract}
 
When a measurement is carried out on one of the entangled parties, the second party can extract work owing to the reduction in its entropy. Here we inquire the amount of work/energy corresponding to the symmetrization entanglement of identical particles (bosons) in a condensate. One measures the quantum state of a particular boson which can be performed only under some certain conditions. We learn that the extracted work comes out to be the \textit{complete} thermodynamical energy present in the condensate.
%We learn that the work extracted by the remaining part of the condensate is equal to the excitation energy of the measured boson times the thermodynamical probability of being in the excited state, i.e., $\hbar \omega_{eg} \times \exp(-\hbar \omega_{eg}/k_BT)$. 
We study the phenomenon in an interacting Bose-Einstein condensate. Then, we discuss that the results may also have fundamental implications on the pair creation in QED vacuum.
\end{abstract}
\maketitle

%%%%%%%%%%%%%%%%%%%%%%%%%%%%%%%%%%%%%%%%%%%%%%%%%%%%%%%%%%%%%%%%%%%%%%%%%%%%%%%%%%%%%%%%%%%%%%%%%%%%%%%%%%%%%%%%%%%%%%%%%%%%%%%%%%%%%%%%
%%%%%%%%%%%%%%%%%%%%%%%%%%%%%%%%%%%%%%%%%%%%%%%%%%%%%%%%%%%%%%%%%%%%%%%%%%%%%%%%%%%%%%%%%%%%%%%%%%%%%%%%%%%%%%%%%%%%%%%%%%%%%%%%%%%%%%%%
%%%%%%%%%%%%%%%%%%%%%%%%%%%%%%%%%%%%%%%%%%%%%%%%%%%%%%%%%%%%%%%%%%%%%%%%%%%%%%%%%%%%%%%%%%%%%%%%%%%%%%%%%%%%%%%%%%%%%%%%%%%%%%%%%%%%%%%%
\subsection{Introduction}
Entanglement can be utilized as a resource in quantum heat engines~\cite{bresque2021two,josefsson2020double,hewgill2018quantum,zhang2007four,rossnagel2014nanoscale,dillenschneider2009energetics,scully2003extracting} where the efficiency can surpass the classical limits~\cite{kieu2004second}. For example, one can employ an ancilla which utilizes quantum correlations for maximizing the extracted work~\cite{francica2017daemonic}. In addition, measurement backaction ---we study here--- can also extract work in quantum entangled  systems~\cite{brunelli2017detecting,cuzminschi2021extractable}. This can be performed, for instance, in a two-mode entangled system as follows. One of the modes, say the $a$-mode, is confined in an optical cavity. A measurement is carried out on the second mode, $b$-mode, which is entangled with the cavity $a$-mode. It is worth noting that the $b$-mode does not need to be in direct contact with the $a$-mode. It can be in another cavity and/or may have been entangled with the $a$-mode via entanglement swap~\cite{pan1998experimental}.

The entropy of the $a$-mode decreases (to $S_V^{\rm (meas)}$) as a backaction of the measurement in the $b$-mode~\cite{brunelli2017detecting}. Both modes are kept in thermal equilibrium with an environment of temperature $T$. That is, the cavity ($a$-) mode rethermalizes with the bath and assigns a larger entropy $S_V^{\rm (ther)}$. During the expansion of the state, the $a$-mode can perform a work in the amount $W=k_B T (S_V^{\rm (ther)}-S_V^{\rm (meas)})$~\cite{maruyama2005thermodynamical}.

A recent study~\cite{tasginEIWE} went a step forward and showed that the environmental monitoring~(measurement) introduces an observer-independent, nature-assigned, correspondence between entanglement and macroscopic, ordered energy~(i.e., kinetic, directional, energy transferred to a mass). In other words, a given amount of entanglement corresponds to a particular amount of directional energy. This takes place because the pointer-basis~(measurement-basis) of the b-mode is solely comprised of coherent states~\cite{zurek1993coherent,gallis1996emergence,tegmark1994decoherence,wiseman1998maximally,zurek1994decoherence,zurek1995decoherence,paraoanu1999selection,zurek1993preferred} when there exists no measurement apparatus, but the environmental modes perform the monitoring of the $b$-mode.  This is, e.g., when the cavity supporting the $b$-mode is placed in vacuum~\cite{zurek1993coherent}. 

The reduced entropy of the $a$-mode, $S_V^{\rm (meas)}$, is already independent from the outcome of the $b$-mode as long as Gaussian measurements and states are concerned. As the measurement-basis is coherent states ---this already implies a Gaussian measurement--- the measurement strength is also fixed at $\lambda=1$~\cite{fiuravsek2002gaussian,giedke2002characterization}. That is, a given degree of entanglement corresponds to a particular amount of ordered energy. This is a result which is independent from the decision of an intellectual creature, like a human, daemon or an ancilla. We call this phenomenon Environmental-induced work extraction~(EIWE). At low temperatures, e.g., room temperature for optical modes, the extracted work (macroscopic energy) is the degree of the entanglement ($\xi$) times the entire thermal energy of the $a$-mode, $W=\xi(r)\times (\bar{n}\hbar \omega_a)$. For max entanglement, $\xi=1$, all of the disordered, microscopic thermal energy present in the $a$-mode is converted into the ordered energy.

In this paper, we inquire work/energy equivalence for the exchange-symmetrization entanglement in an interacting Bose-Einstein condensate~(BEC) made up of ($N+1$) indistinguishable bosons. We wonder the amount of the extracted work when a measurement is carried out on one of the bosonic particles. Though looks simple in words, there are certain conditions for being able to perform a measurement on a single boson (even if it is a random one) in an interacting BEC. We first study these conditions~\cite{stamper1999excitation,stenger1999bragg,andersen2006quantized,tasgin2017many,tacsgin2011creation} in Sec.~\ref{sec:recoil_regimes}. Then, we calculate the work associated with the symmetrization (entanglement) of the measured single boson in Sec.~\ref{sec:work}. When the boson is recoiled and detected in the excited state (of energy $\omega_{eg}$), entropy of the remaining $N$-boson condensate decreases. They do work when they rethermalize with the environment at temperature $T$, e.g., by pushing a board/piston which lies in the condensate region~\cite{brunelli2017detecting,cuzminschi2021extractable}.

The disordered, microscopic thermal energy of the condensate is converted into macroscopic, ordered kinetic energy of the board. At low temperatures, the extracted work (converted energy) comes out to be equal to the excitation energy $\omega_{eg}$ times the thermodynamical probability~($x$) for the condensate to be observed in the excited state, i.e., $W=x \hbar \omega_{eg}$. We note that this is already the entire thermal energy of the condensate either after the rethermalization or before the measurement. In Sec.~\ref{sec:comparison}, we compare the extracted work with the one for the EIWE~\cite{tasginEIWE}. Sec.~\ref{sec:cyclic} examines the energy conservation in a cyclic work extraction process.

Next, in Sec.~\ref{sec:QED}, we consider implications of the results on the physics of electron-positron~(${\rm e}^-$-${\rm e}^+$) quasiparticle pairs created in strong, but subcritical~($E<E_{\rm crt}$) periodic electric fields~\cite{blaschke2006pair,schmidt1998quantum,blaschke2013properties,blaschke2009dynamical}. Here, an (${\rm e}^-$-${\rm e}^+$) quasiparticle pair~\footnote{After the recoil it becomes distinguishable.} stands for the single boson excitation which is distributed symmetrically over all ($N+1$) condensate particles. We discuss the disordered-to-ordered energy conversion in the QED vacuum. Sec.~\ref{sec:summary} contains a brief summary and our conclusions.

%%%%%%%%%%%%%%%%%%%%%%%%%%%%%%%%%%%%%%%%%%%%%%%%%%%%%%%%%%%%%%%%%%%%%%%%%%%%%%%%%%%%%%%%%%%%%%%%%%%%%%%%%%%%%%%%%%%%%%%%%%%%%%%%%%%%%%%%
%%%%%%%%%%%%%%%%%%%%%%%%%%%%%%%%%%%%%%%%%%%%%%%%%%%%%%%%%%%%%%%%%%%%%%%%%%%%%%%%%%%%%%%%%%%%%%%%%%%%%%%%%%%%%%%%%%%%%%%%%%%%%%%%%%%%%%%%
%%%%%%%%%%%%%%%%%%%%%%%%%%%%%%%%%%%%%%%%%%%%%%%%%%%%%%%%%%%%%%%%%%%%%%%%%%%%%%%%%%%%%%%%%%%%%%%%%%%%%%%%%%%%%%%%%%%%%%%%%%%%%%%%%%%%%%%%
\subsection{Recoil regimes of an interacting BEC} \label{sec:recoil_regimes}

Experiments~\cite{stamper1999excitation,stenger1999bragg,andersen2006quantized} and theoretical~\cite{tasgin2017many,tacsgin2011creation,das2016collectively} studies show that responses of an interacting condensate are quite different in the following two regimes. (i) BEC responds to an excitation ``collectively'' when the energy transferred to a boson (e.g., via recoil $\hbar\omega_{\rm r}$) is lower than the mean particle-particle interaction energy $u_{\rm int}=U_{\rm int}/N$~\footnote{Actually, it is quite intriguing that collisions introduce a many-particle entanglement~(MPE) in BECs~\cite{sorensen2001many,tasgin2017many} while in usual they are responsible for decoherence. The interaction term $\hat{\psi}^\dagger ({\bf r}) \hat{\psi}^\dagger ({\bf r}) \hat{\psi}({\bf r}) \hat{\psi}({\bf r}) \propto \hat{S}_z^2$~\cite{sorensen2001many} comes out to be responsible for a many-particle entangled state~[See Fig.~4 in the supplementary material of Ref.~\cite{tasgin2017many}.]. This MPE can be associated with a reduced uncertainty in the number of particles in the quasiparticle spectrum.}, where $U_{\rm int}=\int |\psi({\bf r})|^4 \: d^3r$. Collective response takes place when $\hbar \omega_{\rm r} < u_{\rm int}$. 
In this regime, particles stay in a many-particle entangled~(MPE) state~\footnote{Please see Fig.~4 and the discussion in the supplementary material of Ref.~\cite{tasgin2017many}. See also Ref.~\cite{sorensen2001many}}. (ii) In the contrary, when $\hbar \omega_{\rm r}>u_{\rm int}$, atoms can recoil individually or in groups. (Please, see in particular page 4, paragraph 2 in Ref.~\cite{stenger1999bragg}; page 1, paragraph 2 in Ref.~\cite{stamper1999excitation}; and page 4, paragraph 1 in Ref.~\cite{andersen2006quantized}). Only in this regime, entanglement of a single boson with the condensate can be broken~\cite{tasgin2017many,sorensen2001many}. 

In a typical experiment with a BEC~\cite{inouye1999superradiant}, mean interaction energy is about $u_{\rm int}$ = $10^5$ Hz. Thus, for instance, rotational excitation energy, $\sim 10^2$ Hz~\cite{pethick_smith_Book}, or some part of the phonon excitation energy~\cite{stamper1999excitation} rely in the (i) collective regime. In the contrary, recoil energies associated with the optical light, $\hbar \omega_r=\hbar^2 k^2/2M \sim 10^6$ Hz, falls into the (ii) individual recoil regime~\cite{andersen2006quantized,stenger1999bragg,inouye1999superradiant}. For instance, the superradiance experiments employing BECs rely in the (ii) regime. Also recoil experiments with higher energy phonons~\cite{stamper1999excitation} rely in this regime.

Therefore, one needs to operate the measurement setup in the (ii) regime, $\hbar \omega_{\rm r}$ sufficiently larger than $u_{\rm int}$, in order to address an individual (or a group of) boson(s). This, for example, can be performed by recoiling atoms with single-photons in the measurement setup. Information (measurement) on the state of the recoiled boson can be obtained~(carried out) by monitoring, e.g., the energy/momentum features of the scattered photon. This way, the recoiled boson can be measured either in its ground state $|g\rangle$ or in a higher energy $\omega_{eg}$ (e.g., momentum) state $|e\rangle$. More explicitly, the incident single-photon can recoil a boson either from its ground state or from its excited state. 

We remark that the condition for the recoil of an individual boson is $\hbar \omega_r> u_{\rm int}$. It is not $\hbar \omega_{eg} > u_{\rm int}$. The excited energy of the boson ---for instance $\hbar \omega_{eg}=p^2/2M$ with $p=n\: 2\pi/L$ and $L$ condensate length--- can be can be much smaller than $\hbar\omega_r$ and $u_{\rm int}$.
  
At this step, we need to stress that we are aware of the challenges of the setup we describe above. However, our actual aim here is to study the fundamental consequences of such a measurement regarding the thermal-to-ordered energy conversion. Our actual aim here is not to come up with easy to do experiments.

%%%%%%%%%%%%%%%%%%%%%%%%%%%%%%%%%%%%%%%%%%%%%%%%%%%%%%%%%%%%%%%%%%%%%%%%%%%%%%%%%%%%%%%%%%%%%%%%%%%%%%%%%%%%%%%%%%%%%%%%%%%%%%%%%%%%%%%%
%%%%%%%%%%%%%%%%%%%%%%%%%%%%%%%%%%%%%%%%%%%%%%%%%%%%%%%%%%%%%%%%%%%%%%%%%%%%%%%%%%%%%%%%%%%%%%%%%%%%%%%%%%%%%%%%%%%%%%%%%%%%%%%%%%%%%%%%
%%%%%%%%%%%%%%%%%%%%%%%%%%%%%%%%%%%%%%%%%%%%%%%%%%%%%%%%%%%%%%%%%%%%%%%%%%%%%%%%%%%%%%%%%%%%%%%%%%%%%%%%%%%%%%%%%%%%%%%%%%%%%%%%%%%%%%%%
\subsection{The measurement-induced work} \label{sec:work}

Having outlined the baselines for performing a measurement on identical bosons, we now calculate the work/energy equivalence of the symmetrization entanglement. We consider a condensate consisting of $(N+1)$ number of identical bosons. For the sake of simplicity, we study the low temperature limit. So that, energy of the first excited state remains quite above the thermal fluctuations, i.e., $k_BT \ll \hbar \omega_{eg}$. Here, $\omega_{eg}$ is the level-spacing between the ground and the lowermost excited state. In a BEC, lower-lying energies are associated with the motional degrees of the bosons which are determined by the harmonic trap frequency~\cite{lewenstein1994quantum}, i.e., the depth of optical trap, or the spatial dimension (length=$L$) of the BEC, i.e., $\hbar \omega_{eg}=p^2/2M$ and $p=n\, h/L$~\cite{moore1999quantum}. The level-spacing can be adjusted via the depth of the trap or the dimensions of the condensate.
 
At low temperatures only the first excited state of the condensate
\begin{eqnarray}
|e\rangle_{\scriptscriptstyle N+1} =\big( & |e_1,g_2,...,g_{\scriptscriptstyle N+1} \rangle + |g_1,e_2,...,g_{ \scriptscriptstyle N+1} \rangle + \ldots  \nonumber
\\
& + |g_1,g_2,...,e_{ \scriptscriptstyle N+1} \rangle \: \big)    /\sqrt{N+1}
\label{e_N+1}
\end{eqnarray}
is occupied. In $|e\rangle_{N+1}$, excitation $| e_i\rangle$ is distributed symmetrically among the $N+1$ bosons. For a ``randomly'' chosen boson (the one that recoils in the measurement setup), the same state, Eq.~(\ref{e_N+1}), can be written (still before the measurement) as
\begin{eqnarray}
|e\rangle_{\scriptscriptstyle N+1} = \big[ && \left( |e_1,..,g_{\scriptscriptstyle N} \rangle + \ldots + |g_1,...,e_{\scriptscriptstyle N}\rangle \right) \: | g \rangle_m \nonumber
\\
 &+& |g_1,...,g_{\scriptscriptstyle N}\rangle |e\rangle_t  \quad  \big]/\sqrt{N+1}
 \label{excited_N+1}
\end{eqnarray}
\begin{eqnarray}
=\frac{\sqrt{N}}{\sqrt{N+1}} \: |1\rangle_{\scriptscriptstyle N} \: |g\rangle_{m} + \frac{1}{\sqrt{N+1}} \: |0\rangle_{\scriptscriptstyle N} \: |e\rangle_m,
 \label{excited_N+1_v2}
\end{eqnarray}
where subscript $m$ enumerates the  recoiled boson whose state is measured. This expression is the same for any randomly chosen boson. Here, 
\begin{eqnarray}
&&|1\rangle_{\scriptscriptstyle N}=\left( |e_1,...,g_{\scriptscriptstyle N} \rangle + \ldots + |g_1,...,e_{\scriptscriptstyle N}\rangle \right)/\sqrt{N},
\\
&&|0\rangle_{\scriptscriptstyle N}= | g_1,g_2,...,g_{\scriptscriptstyle N}\rangle
\end{eqnarray}
stand for the normalized $N$-particle Dicke states~\cite{mandel1995optical} where excitation is distributed symmetrically to $N$ bosons. In the $|0\rangle_N$ Dicke state, all of the particles are in the ground state. 
 Similarly, $|e\rangle_{\scriptscriptstyle N+1}=|1\rangle_{\scriptscriptstyle N+1}$ is the $(N+1)$-particle Dicke state. The ground state of the $(N+1)$ system before the measurement is
\begin{equation}
|g\rangle_{\scriptscriptstyle N+1} = |g_1,g_2,....,g_{\scriptscriptstyle N},g_{\scriptscriptstyle N+1}\rangle,
\end{equation}
which can also be represented as $|0\rangle_{\scriptscriptstyle N+1}$. At low temperatures, one can ignore the excited states other than the lowermost one. The thermodynamical probability for the condensate to occupy the excited state behaves as $\propto e^{-\hbar\omega_{eg}/k_BT}$ since one can ignore the chemical potential at low temperatures~\cite{cowan2019chemical}. Thus, the density matrix of the condensate before the measurement can be written as
\begin{eqnarray}
\hat{\rho} \simeq \frac{1}{P_{\rm tot}}  \left( |g\rangle_{\scriptscriptstyle N+1}   \langle g|  +  e^{-\hbar\omega_{eg}/k_B T}  |e\rangle_{\scriptscriptstyle N+1}  \langle e|
 \right),
 \label{rho_before}
\end{eqnarray}
with $P_{\rm tot}=1+e^{-\hbar\omega_{eg}/k_BT}$. We ignore the higher order excitations because $x^2=e^{-\hbar2\omega_{eg}/k_BT}$ is orders of magnitude smaller compared to $x$ and $1$. For simplicity, we also assume that there is no degeneracy in the single-particle excited state $|e\rangle$. From Eq.~(\ref{rho_before}), one can observe that probability for the condensate to observed in the collective excited state is $x=e^{-\hbar\omega_{eg}/k_BT}/P_{\rm tot}\simeq e^{-\hbar\omega_{eg}/k_BT}$. This means that for the repeated experiments/measurements recoil takes place mainly from the ground state.

%One can calculate the von Neumann entropy ($S_V=-{\rm Tr(\hat{\rho} \ln\hat{\rho})}$) of the $N$-particle system (let us abbreviate this system as $A$) before the measurement  by tracing out the $m$th boson. The partial traced density matrix
%\begin{equation}
%\hat{\rho}_A = \frac{1}{P_{\rm tot}}  \left[  \left(1+\frac{x}{N+1}\right) |0\rangle_{\scriptscriptstyle N}   \langle 0|  +  x\left(1+\frac{N}{N+1}\right)  |1\rangle_{\scriptscriptstyle N}  \langle 1|
% \right]
%\end{equation}
%yields the entropy
%\begin{equation}
%S_A^{\rm (before )} \simeq -\frac{xN}{(N+1)(1+x)} \left[  \ln x+\ln\left( \frac{N}{N+1} \right) -\ln(1+x) \right].
%\label{S_Abefore}
%\end{equation}
%The second and the third terms in Eq.~(\ref{S_Abefore}), $\ln\left( \frac{N}{N+1} \right) \ll 1$ and $\ln(1+x)\simeq x \lll 1$, respectively, are both much smaller than $1$. However, the first term $\ln x = -\frac{\hbar\omega_{eg}}{k_B T} \gg 1$ is a large number in magnitude. Thus, the entropy of the not-measured system is
%\begin{equation}
%S_A^{\rm (before )} \simeq x \: \frac{\hbar\omega_{eg}}{k_BT}
%\label{S_Abefore_simpl}
%\end{equation}
%at low $T$.

{\it After the measurement.}--- If the recoiled ($m$th) boson is measured in the excited state, the density matrix of the $N$-particle system becomes
\begin{equation}
\hat{\rho}_A^{\rm (meas)} = |0\rangle_{\scriptscriptstyle N} \langle0|
\end{equation}
after the measurement. Then, the entropy of the $N$-particle system vanishes
\begin{equation}
S_A^{\rm (meas)} = 0.
\end{equation}

{\it Rethermalization.}--- After the measurement, the entanglement between the $N$-particle ($A$-) system and the $m$th boson is broken. The $N$-particle system is still in contact with the environment of temperature $T$, and it rethermalizes. After the rethermalization, density matrix of the $A$ system becomes
\begin{equation}
\hat{\rho}_A^{\rm (ther)} = \frac{1}{P_{\rm tot}} \left[  |0\rangle_{\scriptscriptstyle N} \langle 0| + x |1\rangle_{\scriptscriptstyle N} \langle 1 |    \right].
\end{equation}
Thus, the entropy of the $N$-particle, $A$ system increases back to
\begin{equation}
S_A^{\rm (ther)} = \frac{1}{P_{\rm tot}} \left[  -\ln(1+x) + x\ln x -x\ln(1+x)  \right].
\label{SA_ther}
\end{equation}
The three terms in Eq.~(\ref{SA_ther}) are approximately $\simeq -x$, $\simeq x\frac{\hbar\omega_{eg}}{k_BT}$, and $\simeq x^2$, respectively. In the low $T$ regime, $\frac{\hbar\omega_{eg}}{k_BT} \gg 1$ and it dominates the entropy
\begin{equation}
S_A^{\rm (ther)} \simeq x  \: \frac{\hbar\omega_{eg}}{k_B T}.
\end{equation}

Thus, during the rethremalization one extracts the work
\begin{equation}
W=k_BT \left( S_A^{\rm (ther)} - S_A^{\rm (meas)} \right) = x\: \hbar\omega_{eg}  .
\label{work}
\end{equation}
from the $N$-particle $A$ system.

The extracted work~(\ref{work}) is subject to the measurement of a randomly recoiled boson in the excited state. In case the $m$th boson is measured in the ground state, the entropy change is negligibly small, i.e., order $\sim x^2$.

One should note that $x\hbar \omega_{eg}$ is already the ``entire'' thermodynamical energy of the $N$-particle system. That is, this is already the maximum~(max) thermal energy extractable from this system.
Therefore, we learn that ``entire thermal'' (disordered, microscopic) energy of the system is converted into ordered, directional macroscopic energy when the recoiled boson is detected in the excited state of energy $\hbar\omega_{eg}$. We kindly note that the thermodynamical energy~(\ref{work}) is a result which involves classical probabilities.

This result possibly has important implications regarding the pair-creation process in quantum field theory. This will be discussed in Sec.~\ref{sec:QED}. Before that, we compare the result~(\ref{work}) with other systems.

%%%%%%%%%%%%%%%%%%%%%%%%%%%%%%%%%%%%%%%%%%%%%%%%%%%%%%%%%%%%%%%%%%%%%%%%%%%%%%%%%%%%%%%%%%%%%%%%%%%%%%%%%%%%%%%%%%%%%%%%%%%%%%%%%%%%%%%%
%%%%%%%%%%%%%%%%%%%%%%%%%%%%%%%%%%%%%%%%%%%%%%%%%%%%%%%%%%%%%%%%%%%%%%%%%%%%%%%%%%%%%%%%%%%%%%%%%%%%%%%%%%%%%%%%%%%%%%%%%%%%%%%%%%%%%%%%
%%%%%%%%%%%%%%%%%%%%%%%%%%%%%%%%%%%%%%%%%%%%%%%%%%%%%%%%%%%%%%%%%%%%%%%%%%%%%%%%%%%%%%%%%%%%%%%%%%%%%%%%%%%%%%%%%%%%%%%%%%%%%%%%%%%%%%%%
\subsection{Comparison with environment-induced work extraction} \label{sec:comparison}

We now compare the results for the two phenomena: work/energy corresponding to (i) the symmetrization entanglement in an ensemble, $W=x\, \hbar\omega_{eg}$, and (ii) EIWE, $W=\xi(r)\times (\bar{n}\hbar \omega_a)$~\cite{tasginEIWE} where entanglement is between the optical modes $a$ and $b$. The main differences between the two are the followings. In EIWE, work is extracted by itself via monitoring of the environment. No intellectual being exists unlike the extraction from an ensemble. 
Moreover, the extracted amount does not depend on the measurement outcome of the $b$-mode while in the case of an ensemble it does.

The two phenomena are similar in the following way. Entire thermodynamical energy is converted into work for max entanglement. In EIWE, $\xi(r)$ stands for the degree of the entanglement between the two modes. It increases from $0$ to $1$ (max value) with the increasing degree of entanglement. For $\xi=1$, all thermal energy, $\bar{n}\hbar\omega_a$, is converted into work. Now taking a look at the first excited state of the condensate, Eq.~(\ref{excited_N+1}), one can realize that it is a maximally entangled state with respect to any one of the recoiled~(measured) particles. Again, extracted work is the complete thermal energy $x\hbar\omega_{eg}$.

We can even provide a third example into the comparison. The extracted work for the max entangled two-mode state (iii) $|e\rangle=(|1,0\rangle + |0,1\rangle)/\sqrt{2}$ also comes out as $(\bar{n}\hbar \omega_a)$~\cite{tasginEIWE}. This is similarly the entire thermodynamical energy of the system. This result, similar to the (i) symmetrization entanglement, is also subject to the measurement of the $b$-mode in the excited state $|1\rangle_b$. 

Therefore, the comparison among (i)-(iii) tells us the following. In case of max entanglement, the extracted work/energy becomes the entire thermodynamical energy at low temperatures. What ever present in the mode/system is converted into the ordered macroscopic energy. This is simply because max entanglement collapses the other party in a deterministic state with zero entropy. We remind that the calculated work in each case involves thermodynamical (classical) probabilities.

%%%%%%%%%%%%%%%%%%%%%%%%%%%%%%%%%%%%%%%%%%%%%%%%%%%%%%%%%%%%%%%%%%%%%%%%%%%%%%%%%%%%%%%%%%%%%%%%%%%%%%%%%%%%%%%%%%%%%%%%%%%%%%%%%%%%%%%%
%%%%%%%%%%%%%%%%%%%%%%%%%%%%%%%%%%%%%%%%%%%%%%%%%%%%%%%%%%%%%%%%%%%%%%%%%%%%%%%%%%%%%%%%%%%%%%%%%%%%%%%%%%%%%%%%%%%%%%%%%%%%%%%%%%%%%%%%
%%%%%%%%%%%%%%%%%%%%%%%%%%%%%%%%%%%%%%%%%%%%%%%%%%%%%%%%%%%%%%%%%%%%%%%%%%%%%%%%%%%%%%%%%%%%%%%%%%%%%%%%%%%%%%%%%%%%%%%%%%%%%%%%%%%%%%%%
\subsection{Energy conservation in a cyclic extraction process} \label{sec:cyclic}

In this section, we investigate the situation when work extraction is continued by recoil/measurement of new bosons. We try to gain a better understanding on the amount of energy that the condensate draws from the environment and we examine what does ``entire thermal energy'' correspond in such a cyclic process. Taking into account the energy conservation, the energy drawn turns out to be $\hbar\omega_{eg}$ in the mid-steps.

Before the measurement is performed, that is before the $m$th boson is observed in the excited state, the entire (mean) thermal energy of the condensate [$(N+1)$ atoms] was $x\hbar\omega_{eg}$. This is merely a mean involving averages over classical thermodynamical (grand-canonical) probabilities. Once the $m$th boson is measured in the excited state, however, one learns that the energy of the ensemble was/is $\hbar\omega_{eg}$. Now, it is not an average. After the remaining part of the condensate rethermalizes, the mean energy of the condensate becomes again $x\hbar\omega_{eg}$. The remaining part of the condensate (we mean the $N$-particle system $A$) draws energy from the environment and while doing this it does work on a board/piston. 

We can continue this cycle, e.g., one more time. The density matrix of the $N$-particle $A$ system is the same with the one in Eq.~(\ref{rho_before}) except that now it consists $N$ boson. We remark that in the recoil/measurement of the $m$th boson, entanglement between the $m$th boson and the $N$-particle $A$ system is broken. The remaining part of the condensate ($A$) behaves still collectively and it is in the $N$-particle version of the state~(\ref{rho_before}). Thus, if we recoil/measure the $m_2$th boson in the excited state, entropy of the remaining part of the condensate [of $(N-1)$ bosons] vanishes. This means that it can extract the same amount of work one more time by rethermalizing~\footnote{We assume that $N$ is very large in the calculations for the sake of simplicity.We also note that the first excited energy of the condensate always possesses a single excitation which is distributed symmetrically among the number of constituted bosons. We tell this because after the rethermalization in the first cycle, the measurements can recoil ground state atoms till the recoil with an excited state boson takes place. Naturally, the probability for finding a boson in the ground state is accordingly higher.}. One more time, in the middle step the system is determined to have the $\hbar\omega_{eg}$ energy. One can continue the cycle as long as she/he desires.

Thus, the measurements in the middle steps witness that the condensate possesses $\hbar\omega_{eg}$ energy. Thus, one can understand that during the rethermalization processes of the cycle ---in which work extraction is performed--- condensate actually draws an $\hbar\omega_{eg}$ energy from the environment. The $x\hbar\omega_{eg}$ is only a mean over thermodynamical probabilities. This is the actual reason for us ti signify that the extracted work (the amount of disordered-to-ordered energy conversion) as the ``entire thermal energy'' present in the condensate/mode throughout the paper. 
This finding has important implications in pair (a boson together) creation process in the QED vacuum, discussed in the following section.

{\it Quasiparticle excitations.}--- Before discussing the QED implementations, we would like to point out the following important issue. The collective excitations of an ensemble are often referred as quasiparticle excitations~\cite{jurcevic2014quasiparticle,sachdev2000quantum,emary2003chaos}. They can be represented by bosonic annihilation/creation operators, e.g., $\hat{c}$ and $\hat{c}^\dagger$~\cite{emary2003chaos}. In this representation, the symmetrized single-particle excited state~Eq.~(\ref{e_N+1}) can be equivalently represented as $|1\rangle$. Therefore, in this picture, we can also tell that when the realization of the quasiparticle excitation $|1\rangle_{\scriptscriptstyle N+1}$ as an independent particle takes place, the ensemble extracts the max possible work.

%%%%%%%%%%%%%%%%%%%%%%%%%%%%%%%%%%%%%%%%%%%%%%%%%%%%%%%%%%%%%%%%%%%%%%%%%%%%%%%%%%%%%%%%%%%%%%%%%%%%%%%%%%%%%%%%%%%%%%%%%%%%%%%%%%%%%%%%
%%%%%%%%%%%%%%%%%%%%%%%%%%%%%%%%%%%%%%%%%%%%%%%%%%%%%%%%%%%%%%%%%%%%%%%%%%%%%%%%%%%%%%%%%%%%%%%%%%%%%%%%%%%%%%%%%%%%%%%%%%%%%%%%%%%%%%%%
%%%%%%%%%%%%%%%%%%%%%%%%%%%%%%%%%%%%%%%%%%%%%%%%%%%%%%%%%%%%%%%%%%%%%%%%%%%%%%%%%%%%%%%%%%%%%%%%%%%%%%%%%%%%%%%%%%%%%%%%%%%%%%%%%%%%%%%%
\subsection{Implementations with pair creation within vacuum fluctuations} \label{sec:QED}

In the previous section, we realized that an ensemble of bosons converts all of the thermodynamical energy into macroscopic energy upon the measurement of a recoiled boson in the excited state. This takes place because the recoiled atom is maximally entangled with the remaining condensate. Similarly, all of the thermal energy present in a photonic mode is also converted into work if it is maximally entangled with another mode. In the previous section, we also considered a cyclic extraction process for the condensate. We realized that energy drawn from the environment comes out to be $\hbar\omega_{eg}$, i.e., the energy of the excited state in which the recoiled boson was measured. This follows from the conservation of energy. 

In this section, we inquire if conversion of disordered energy into ordered energy has something to do with pair generation in the QED vacuum~\cite{schwinger1951gauge,gelis2016schwinger}. We study an analogy between the recoil/measurement of a boson in a condensate and realization of, e.g., an electron-positron pair in the QED vacuum under subcritical dynamical electric field~\cite{blaschke2006pair,schmidt1998quantum,blaschke2013properties,blaschke2009dynamical}. We examine the thermodynamical consequences of such an analogy.

%A pair of fermionic particles behaves as a bosonic quasiparticle excitation similar to the ones we study in our condensate. Their symmetrization needs to be taken into account []. 

Energy of the vacuum, on which quasiparticles are excited, possesses the largest known discrepancy lying between $10^{-9}$ Joules/${\rm m}^3$ (calculated from the upper limit of the cosmological constant) and $10^{113}$ Joules/${\rm m}^3$ predicted by QED~\cite{QED_vacuum}. In quantum field theory~(QFT), particle pairs are excited in this vacuum~\cite{peskin2018introduction}. Studies on BECs~\cite{stamper1999excitation,stenger1999bragg,andersen2006quantized,tasgin2017many,sorensen2001many,tacsgin2011creation,das2016collectively} demonstrate that a quasiparticle [i.e., symmetrized superposition state where one or more particle is in the excited state, e.g., Eq.~(\ref{e_N+1})] behaves as a ``particle'' (independently) when the recoil energy delivered to the particle exceeds a critical value ---the mean interaction energy $u_{\rm int}$. Otherwise, the condensate responds a disturbance only collectively~\cite{stamper1999excitation,stenger1999bragg,andersen2006quantized} as it remains in the many-particle entangled state~\cite{tasgin2017many,sorensen2001many}. Thus, one of the particles cannot simply be broken away from the condensate.

Let us consider pair creation processes in vacuum, such as electron-positron~\cite{schwinger1951gauge,alkofer2001pair} or quark-antiquark generation~\cite{casher1979chromoelectric,sonner2013holographic,jensen2013holographic} under strong electric or chromo-electric fields, respectively. In particular, electron-positron generation takes place above the critical electric field $E>E_{\rm crt}$~\cite{schwinger1951gauge,gelis2016schwinger} where the field becomes strong enough to transfer an amount of $2m_e c^2$ energy in order to generate the pair. Here, we are interested in another regime, when the strong electric field lies sufficiently below $E_{\rm crt}$, i.e., subcritical field. In this regime, the strong electric field introduces plasma of electron-positron {\it quasiparticle} pairs in the QED vacuum~\cite{blaschke2006pair,schmidt1998quantum,blaschke2013properties,blaschke2009dynamical}~\footnote{These results follow from the nonperturbative solutions of the Dirac equation~\cite{schmidt1998quantum}. The matter field is quantized and electric field is treated classically. For instance, in Ref.~\cite{blaschke2006pair} electron-positron quasiparticle pairs for the two counter-propagating pulses are investigated. The dynamics of the pair creation is investigated, i.e., dynamical QED, in periodic time-dependent electric fields in difference to a constant uniform electric field investigated earlier by Schwinger~\cite{schwinger1951gauge}.}. Quasi-energy spectrum tells us that these quasiparticle excitations are not on the mass shell because they do not satisfy the energy momentum relation~\cite{blaschke2013properties}~\footnote{More explicitly, these virtual particles do not satisfy $E^2-p^2c^2=(mc^2)^2$.}.

 Here, we analogize the condensate with the fluctuating QED vacuum under the subcritical electric field~\cite{blaschke2006pair,schmidt1998quantum,blaschke2013properties,blaschke2009dynamical}. In the condensate, the excitation is distributed symmetrically over ($N+1$) bosons. Actually, this is a quasiparticle~\cite{jurcevic2014quasiparticle,sachdev2000quantum,emary2003chaos} which can be represented as $|N+1,1\rangle$~\cite{mandel1995optical}, or simply as $|1\rangle_{\scriptscriptstyle N+1}$. Thus, we analogize this quasiparticle excitation with an electron-positron quasiparticle pair studied in Refs.~\cite{blaschke2006pair,schmidt1998quantum,blaschke2013properties,blaschke2009dynamical}. Please note that in our condensate and QED vacuum there could be more than one quasiparticle excitation, e.g., $|N+1,2\rangle$, which should be distributed symmetrically over all of the condensate atoms. Within this analogy, the quantum state of the electron-positron quasiparticle pair is distributed among many parties interacting with each other~\footnote{At this point, we better underline that our analogy fundamentally differs from the studies which employ BECs as simulators of, e.g., nonlinear Dirac equations~\cite{gerritsma2010quantum,haddad2015nonlinear}. These studies, a well-established area now, map the interaction dynamics among already-generated-particles onto the nonlinear interaction term present in the BEC hamiltonian. They aim to simulate the high-energy processes/interactions employing the experiments with low-energy cold atoms. Here, we perform a different analogy. We analogize the collective response of a BEC with the one belonging to the QED vacuum. In a BEC, particle-particle interactions are responsible for the collective behavior where a single boson cannot be observed/measured unless it is recoiled by $\omega_r < u_{\rm int}$. That is, we analogize the interparticle interactions with the ones taking place in the QED vacuum, i.e., not among the already generated (recoiled) pairs.}. 
Before going on further with the analogy, we remind that the condensate responds only collectively for $\hbar\omega_{r}<u_{\rm int}$. That is, one cannot recoil/observe a single boson unless such a recoil energy is delivered. This is because, the boson/condensate stays in a many-particle entangled state. We analogize this behavior with the response of the QED vacuum where subcritical electric field is introduced. An electron-positron quasiparticle pair needs to be recoiled in order the quasiparticle to become a real pair (on the mass shell)~\cite{blaschke2013properties}~\footnote{The recoil can be delivered by another excitation/field.}. Now, after the recoil, the exchange symmetrization entanglement between the pair and the condensate (QED vacuum) is broken.

Up to here, the analogy provided only a parallel description for the pair creation. However, our calculations with a BEC also tells us the following. Upon the recoil and observation of the single boson in the excited state (creation/observation of the pair in the analogy), the remaining part of the condensate (QED vacuum in the analogy) draws  energy from the environment~\footnote{We have in mind astronomically large masses of different vacua, i.e., environment. The QED vacuum, within which our pair is generated, is one of these vacua and it interacts with this environment.}. The extracted energy is the complete thermodynamical energy present in the condensate, $x\hbar\omega_{eg}$, either before the measurement or after the rethermalization. Moreover, the discussion given in the previous section tells us that the disordered energy drawn from the environment and converted into work should be $\hbar\omega_{eg}$ in a cyclic extraction process. This is when more than one boson (pair) are recoiled (created).

Thus, one can tell the following. When a pair is generated, the vacuum generating the pair draws an $x\,\hbar\omega_{eg}$ amount of work from the environment. But this is a thermodynamical mean only. The conservation of energy in a cyclic process tells us that in each observation of a real pair, the QED vacuum draws an $\hbar\omega_{eg}$ amount of energy from the environment. What is more important is that amount of disordered thermal energy present in the environment is converted into ordered, macroscopic, directional energy (work) when the pair is created/observed. Thus, this stands as something like a conservation relation. In brief, QED vacuum pays out from its inherited symmetrization entanglement for generating ordered masses of particles.

%%%%%%%%%%%%%%%%%%%%%%%%%%%%%%%%%%%%%%%%%%%%%%%%%%%%%%%%%%%%%%%%%%%%%%%%%%%%%%%%%%%%%%%%%%%%%%%%%%%%%%%%%%%%%%%%%%%%%%%%%%%%%%%%%%%%%%%%
%%%%%%%%%%%%%%%%%%%%%%%%%%%%%%%%%%%%%%%%%%%%%%%%%%%%%%%%%%%%%%%%%%%%%%%%%%%%%%%%%%%%%%%%%%%%%%%%%%%%%%%%%%%%%%%%%%%%%%%%%%%%%%%%%%%%%%%%
%%%%%%%%%%%%%%%%%%%%%%%%%%%%%%%%%%%%%%%%%%%%%%%%%%%%%%%%%%%%%%%%%%%%%%%%%%%%%%%%%%%%%%%%%%%%%%%%%%%%%%%%%%%%%%%%%%%%%%%%%%%%%%%%%%%%%%%%
\subsection{Summary and Conclusions} \label{sec:summary}

In summary, we calculate the work associated with the symmetrization entanglement. In other words, we calculate the amount of disordered thermal energy which is converted into a macroscopic ordered energy. The ordered energy is associated with the entanglement. First, we briefly mention how one can carry out a measurement on one of the (random) identical particles in a condensate. When one of the particles is delivered a recoil energy which is above the mean-interaction energy, $\hbar\omega_r>u_{\rm int}$, the particle can be recoiled from the condensate. In this regime, symmetrization entanglement of a particle with the condensate can be broken. Otherwise, condensate responds collectively, e.g., one can consider the collective rotations in a condensate which takes place only after each boson receives $\hbar$ amount of angular momentum~\cite{pethick_smith_Book}.

Our results show that the entire thermal energy present in the condensate is converted into  work (ordered energy), when the recoiled boson is measured in the excited state. The full-energy-conversion takes place because the measured particle is maximally entangled with the remaining part of the condensate. Next, we investigate cyclic work extraction from a condensate. That is, after the rethermalization of the condensate in the first cycle another measurement is performed. When a second boson is recoiled/detected, the conservation of energy tells us that the energy drawn from the environment is $\hbar\omega_{eg}$.

Next, we wonder weather such a disordered-to-ordered energy conversion and the conservation-like behavior (mentioned above) have something to with the pair creation from the QED vacuum. QED vacuum where a subcritical electric field ($E$ is strong but sufficiently lower than $E_{\rm crt}$) is introduced can be characterized by electron-positron quasiparticle pairs~\cite{blaschke2006pair,schmidt1998quantum,blaschke2013properties,blaschke2009dynamical}. There is a similarity between the quasiparticle excitation of a BEC, Eq.~(1), and an electron-positron quasiparticle pair. One cannot measure/observe one of the bosons in a BEC or a pair in the QED vacuum unless a recoil is delivered to the boson or pair, respectively. 

The analogy tells us that the QED vacuum, at which pair(s) is(are) realized, converts the entire disordered thermal energy into the ordered one (work) when an electron-positron pair is realized~\footnote{We remind that here we consider a QED vacuum on which subcritical electric field is introduced. It possesses electron-positron quasiparticle pairs~\cite{blaschke2006pair,schmidt1998quantum,blaschke2013properties,blaschke2009dynamical}. }. The QED vacuum draws the disordered energy from the environment, i.e., from masses of other vacua. It uses the inherited symmetrization entanglement for converting the disordered energy into ordered masses of pairs.

Finally, we would like to conclude the paper with a concrete model for the QED vacuum generating the pair(s). A related example could be the crystal vibrations in a crystal, i.e., the sound excitations. In this system, where nearest-neighbor interactions are in the play, the quantum of the quasiparticle excitations is already determined by the interaction strength~\footnote{Hamiltonian in the presence of nearest-neighbor interactions in a crystal of atoms can be written as $\hat{H}=\sum_{s=1}^N \left\{ \frac{p_s^2}{2M} + \frac{1}{2}C(q_{s+1}-q_s)^2 \right\}$ where $q_s$ and $p_s$ are the position and momentum coordinates of the atoms. Employing the quantization algebra~\cite{kittel2018introduction}, the phonon energy quantum is determined as $\omega_k = \left( \frac{2C}{M} \right)^{1/2} \, \left[ 1-\cos(ka) \right]^{1/2}$ where $a$ is the lattice spacing and $k$ is the phonon momentum. Thus, in this case we observe that $C\propto M\omega_k^2$.}.
The two, excitation energy and the mean-interaction energy, are in the same order.

%%%%%%%%%%%%%%%%%%%%%%%%%%%%%%%%%%%%%%%%%%%%%%%%%%%%%%%%%%%%%%%%%%%%%%%%%%%%%%%%%%%%%%%%%%%%%%%%%%%%%%%%%%%%%%%%%%%%%%%%%%%%%%%%%%%%%%%%
%%%%%%%%%%%%%%%%%%%%%%%%%%%%%%%%%%%%%%%%%%%%%%%%%%%%%%%%%%%%%%%%%%%%%%%%%%%%%%%%%%%%%%%%%%%%%%%%%%%%%%%%%%%%%%%%%%%%%%%%%%%%%%%%%%%%%%%%
%%%%%%%%%%%%%%%%%%%%%%%%%%%%%%%%%%%%%%%%%%%%%%%%%%%%%%%%%%%%%%%%%%%%%%%%%%%%%%%%%%%%%%%%%%%%%%%%%%%%%%%%%%%%%%%%%%%%%%%%%%%%%%%%%%%%%%%%
\subsection*{Acknowledgements}

We are grateful to Vural G\"{o}kmen for the motivational support. 
We acknowledge the scientific support of Cemsinan Deliduman and Bayram Tekin on high-energy physics.

\bibliography{bibliography}

\end{document}